\documentclass[runningheads]{llncs}
\usepackage{multirow}
\usepackage{graphicx}
\usepackage{booktabs}

\usepackage[inkscapeformat=png]{svg}

\usepackage{amsmath,amssymb,amsfonts} 
\usepackage{booktabs} 
\begin{document}

\title{Enhanced segmentation of femoral bone metastasis in CT scans of patients using synthetic data generation with 3D diffusion models}

\titlerunning{Metastasis segmentation using synthetic CT data generation}

\author{Emile Saillard\inst{1,2}\and
Aurélie Levillain\inst{3} \and
David Mitton \inst{3}
Jean-Baptiste Pialat \inst{2,4}
Cyrille Confavreux \inst{1,4}
Hélène Follet\inst{1*c}\and
Thomas Grenier\inst{2*}}

%
\authorrunning{E. Saillard et al.}
%
\institute{INSA-Lyon, Université Claude Bernard Lyon 1, CREATIS, UMR5220, 69100 Villeurbanne, France 
\and Université Claude Bernard Lyon 1, INSERM, LYOS UMR 1033, 69008 Lyon, France 
\and Université Claude Bernard Lyon 1, Université Gustave Eiffel, LBMC UMRT9406, 69622, Lyon, France 
\and Hospices Civils de Lyon, France}

\maketitle              
   ~\\
* These authors contributed equally 
\\
c Corresponding author
\\

\textbf{Manuscript Type:} Original Research \\

\textbf{Summary Statement:}\\
\textbf{In this paper, we propose a data synthesis pipeline that generates realistic and diverse metastatic volumes, allowing to train lesions segmentation models more efficiently and obtain better performance than other state-of-the-art models. }\\

\textbf{Key Points:}
\begin{itemize}
    \item The proposed pipeline allows to produce a large number of realistic CT volumes (5675) with highly variable lesions in femur, 
    \item Diffusion models improve the quality of generated CT-volumes when considering lesions segmentation performances of deep learning networks,
    \item Such trained networks allow us to outperform state-of-the-art lesions segmentation approaches.
\end{itemize}

\begin{abstract}~\\
    
\textbf{Purpose:} \\ Bone metastasis have a major impact on the quality of life of patients and they are diverse in terms of size and location, making their segmentation complex.
Manual segmentation is time-consuming, and expert segmentations are subject to operator variability, which makes obtaining accurate and reproducible segmentations of bone metastasis on CT-scans a challenging yet important task to achieve. \\

\textbf{Materials and Methods: }\\
Deep learning methods tackle segmentation tasks efficiently but require large datasets along with expert manual segmentations to generalize on new images.
We propose an automated data synthesis pipeline using 3D Denoising Diffusion Probabilistic Models (DDPM) to enchance the segmentation of femoral metastasis from CT-scan volumes of patients.
We used 29 existing lesions along with 26 healthy femurs to create new realistic synthetic metastatic images, and trained a DDPM to improve the diversity and realism of the simulated volumes. We also investigated the operator variability on manual segmentation.\\

\textbf{Results:} \\ We created 5675 new volumes, then trained 3D U-Net segmentation models on real and synthetic data to compare segmentation performance, and we evaluated the performance of the models depending on the amount of synthetic data used in training.

\textbf{Conclusion:} \\Our results showed that segmentation models trained with synthetic data outperformed those trained on real volumes only, and that those models perform especially well when considering operator variability.

\end{abstract}
%
%
\section{Introduction} 

Bone metastasis are frequent in cancer patients, particularly those with primary prostate or breast cancer \cite{macedo2017}. 
These metastasis often lead to various complications, like pathological fractures, having a significant impact on patients' quality of life. 
Computed tomography (CT) is a widely utilized imaging technique to assess the risk of pathological fractures clinically.

Segmenting bone metastasis poses significant challenges due to the diverse types of metastasis, varying sizes, and lesion locations. 
Manual segmentation of these lesions is particularly time-consuming, reflecting the complexity of the task. 
Moreover, intra- and inter-operator variability in bone metastasis segmentation is substantial, especially for smaller lesions \cite{ataei_evaluation_2021}.
This variability, although dataset-dependant, underscores the complexity of segmentation and affects applications like biomechanical simulations where reproducibility is key and sensitivity to segmentation is important \cite{gardegaront2024inter}.
Models trained for automatic segmentation of femoral bone metastasis using CT data have been documented \cite{ataei_effect_2024,huo_deep_2023,rachmil2024automatic}. 
However, the segmentation quality achieved by these models is insufficient for clinical use \cite{paranavithana_systematic_2023}. 
The lack of public datasets with CT images of femoral metastasis hinders model comparison, as performance depends heavily on the dataset used.
Moreover, the lack of manually annotated data complicates the training process, leading to lower generalizability of trained models due to overfitting.
Data augmentation techniques are commonly applied to prevent overfitting, but these operations may not introduce the variability required for optimal model generalization.
Using synthetic data can also aid in training \cite{fernandez2022can}, leveraging networks such as Generative Adversarial Networks \cite{goodfellow_GAN_2014} or Denoising Diffusion Probabilistic Models (DDPM) \cite{ho_ddpm_2020,kazerouni2023diffusion}. 
However, these networks demand a substantial volume of training data to generate diverse samples.
Pathological CT volumes of bone metastasis with expert manual segmentation are rare and custom methods can be explored to generate synthetic data. 
We propose a novel synthesis pipeline utilizing a tailored algorithm to create realistic pathological images using manually segmented lesions and healthy femurs to improve the segmentation performance of Deep Learning models.
Additionally, we explore the use of 3D diffusion networks to increase the diversity of the synthetic volumes generated.
We produced a substantial amount of synthetic images and evaluated the segmentation performance of 3D U-Nets \cite{ronneberger2015} trained on real and/or synthetic data.
To further validate the use of our models, we investigated operator variability and compared the results with automatic segmentations. 
We share the code, and part of real and synthesized images, with corresponding annotations.

\section{Materials and Methods} 
In this section, we describe the datasets used to train and evaluate the different models.
We outline the different processing steps and model architectures employed for image synthesis and bone metastasis segmentation. This is a prospective study.

\subsection{Data}
The dataset used for metastasis segmentation consists of 29 manually segmented CT-scans of anonymized patients with femoral osteolytic metastasis (National number: 2019-A01202-55): 20 from the MEKANOS cohort ((Hospices civils Lyon, agreement number N. 21\_5467, May 28th 2021)), 9 from the DEMETOS cohort (Hospices civils Lyon, RNIPH, MR-004-21\_5467). Additionally, 9 femoral CT scans of patients from the DEMETOS cohort from 3 different centers were also used to further evaluate segmentation performance and to evaluate operator variability.

\begin{figure}[h]
    \includegraphics[width=\columnwidth, keepaspectratio]{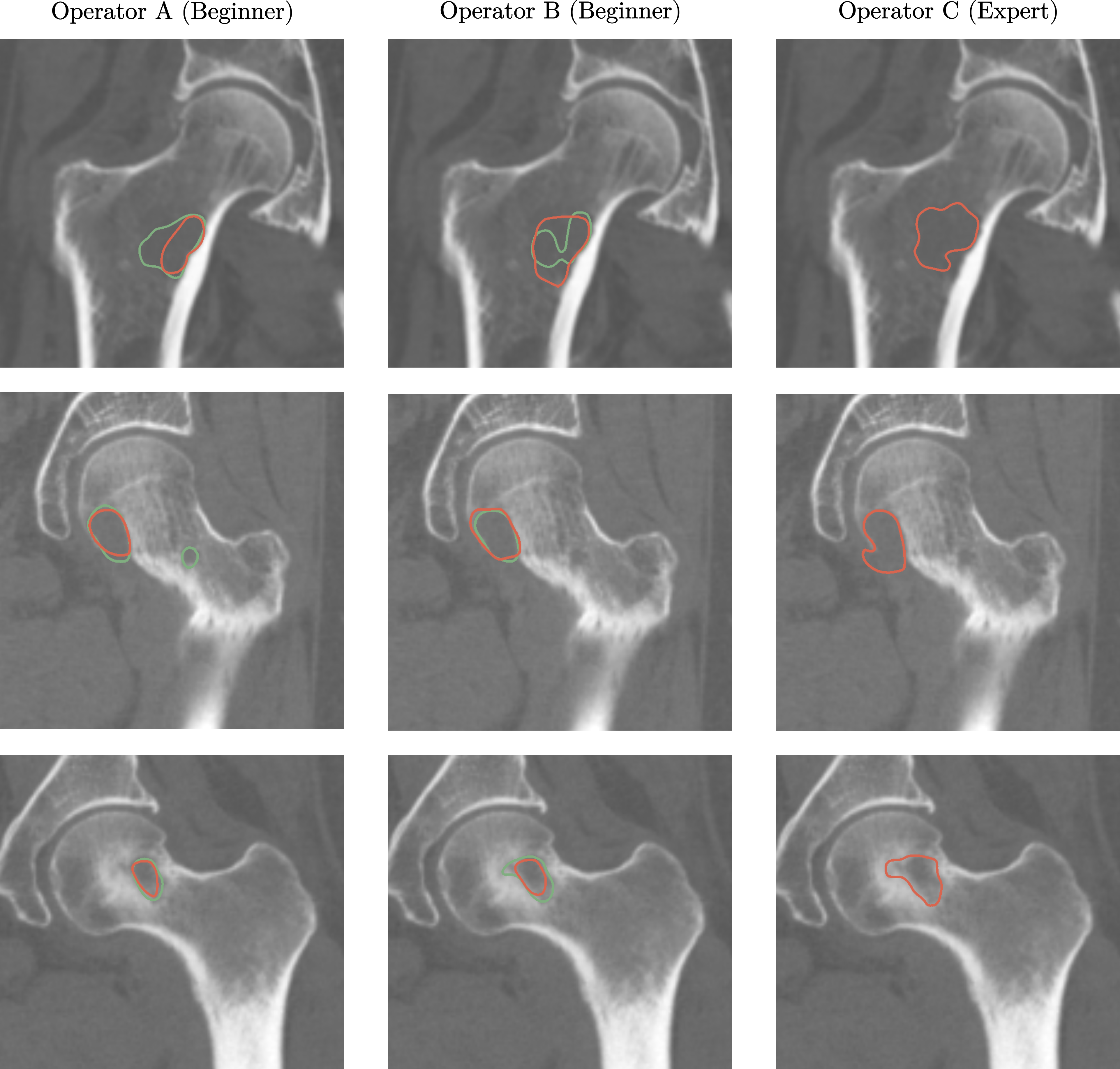}
    \caption{\textsf{Examples of femoral CT scans from three patients with metastasis in coronal view with operator segmentations (first annotation in red and second annotation few months later in green for beginners) }}
    \label{fig:data_operators}
\end{figure}

Examples of femurs from the 3 centers used for this study are displayed in Figure \ref{fig:data_operators}, along with operator segmentations.
The scans from MEKANOS were acquired in clinical routine following a specific procedure (constant table height, quality phantom QA Mindways, 120~kV, 270~mAs, 1 Pitch, Field of view 360~mm and 200~mm, reconstruction: 
standard filter B, $ 512 \times 512 $ matrix, slice thickness 0.7~mm) and with three manufacturers' acquisition systems (General Electric, Philips and Siemens).
The remaining scans were not acquired with a standard protocol and had therefore various area scanned, acquisition systems used as well as voxel sizes.
All femurs have at least one osteolytic metastasis, each manually segmented by expert radiologists.
We only considered lesions bigger than 26 voxels ($16mm^3$), fixed by the size of the smallest real lesions of our dataset.

For the data synthesis task, 29 scans with metasasis as well as 26 additional healthy femurs from the DEMETOS and MEKANOS cohorts were used.
All volumes were resampled to median voxel size (0.85 × 0.85 × 0.85mm) and intensities were standardized.

The synthetic images and their annotations are available at 10.5281/zenodo.13771568.

\subsection{Operator variability} 
We evaluate operator variability on 9 metastatic femurs from 3 different centers.
Two novice operators A and B (one master student and an assistant professor in biomechanics) and an expert radiologist (C) manually segmented the femoral metastasis. 
The DICE score is calculated between the operators to quantify inter-operator variability. 
To assess intra-operator variability, operators A \& B segmented the same metastasis again after three months. 
We also compare manual segmentations to automatic segmentation (model "diffusion + fine-tuning") on those 9 metastatic femurs.

\subsection{Data synthesis}
\subsubsection{Ad-hoc data synthesis} 
\label{ssec:adhoc_data_synthesis}

\begin{figure}[h]
    \includegraphics[width=\columnwidth, keepaspectratio]{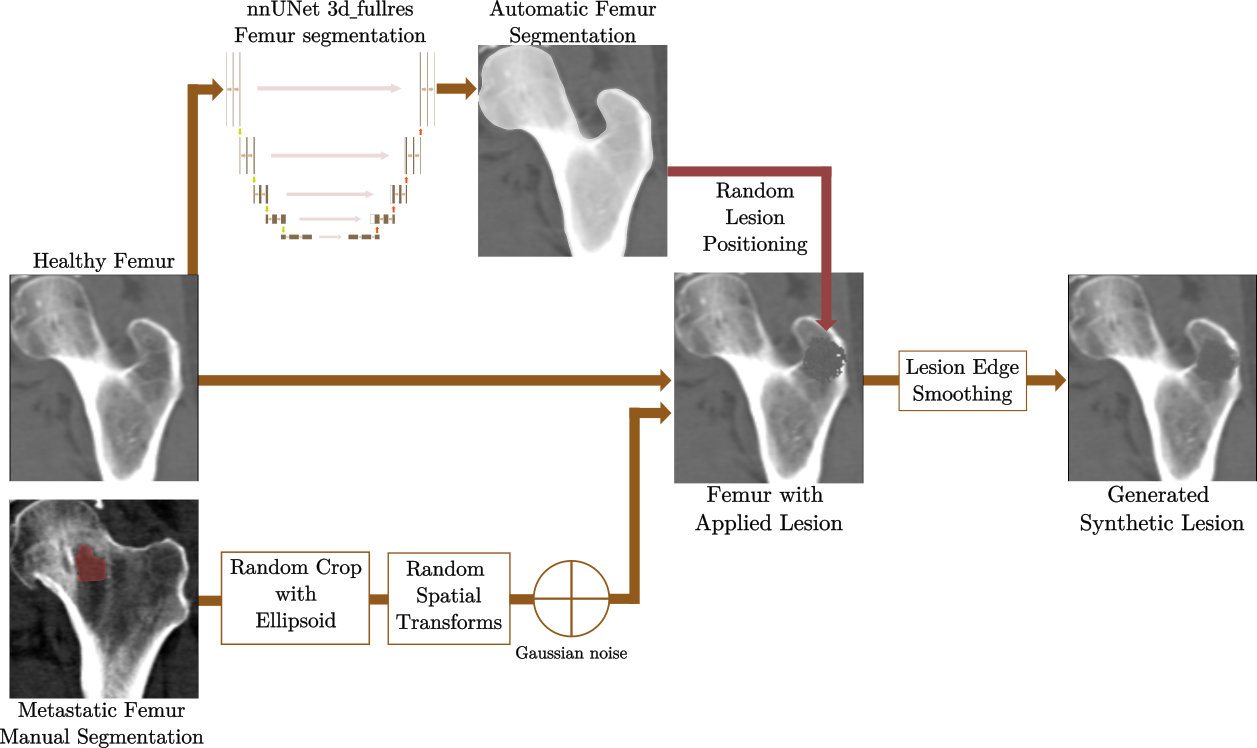}
    \caption{\textsf{Proposed fully automated image synthesis pipeline of CT femoral metastasis used to generate realistic pathological images using existing lesions and healthy femurs.}}
    \label{fig:synthesis_diagram}
\end{figure}

To improve the robustness of our segmentation models, we propose a fully automated image synthesis pipeline that increases the quantity of training data by generating realistic pathological images using existing lesions and healthy femurs, as illustrated in Figure \ref{fig:synthesis_diagram}.

Using the manual segmentation of the metastasis, the lesion is first extracted from the original CT volume and then cropped via a random intersection with an ellipsoid shape, mimicking a realistic lesion shape.
Spatial transforms (rotations \& scaling) are then applied to the lesion.
For greater realism and diversity, the edges of the lesion are smoothed by a mean filter, followed by the addition of Gaussian noise.
The resulting lesions are then applied to the healthy femurs which are automatically segmented as in \cite{saillard2024finite} to constrain the random placement of the extracted lesion inside the femur.

Using this pipeline, we generate 7540 new volumes, which corresponds to 10 applications on the 26 healthy femurs of the lesions extracted from the 29 pathological volumes.
Images where the random positioning produced too small lesions ($\leq 16mm^3$) were removed, decreasing the number of usable volumes to 5675.
In the remainder of this paper, we named this set "Synthetic".
All the volumes of the synthetic sets were then processed using the diffusion schemes proposed in the following section.

\begin{figure}[tp]
\centering
    \includegraphics[width=\columnwidth, keepaspectratio]{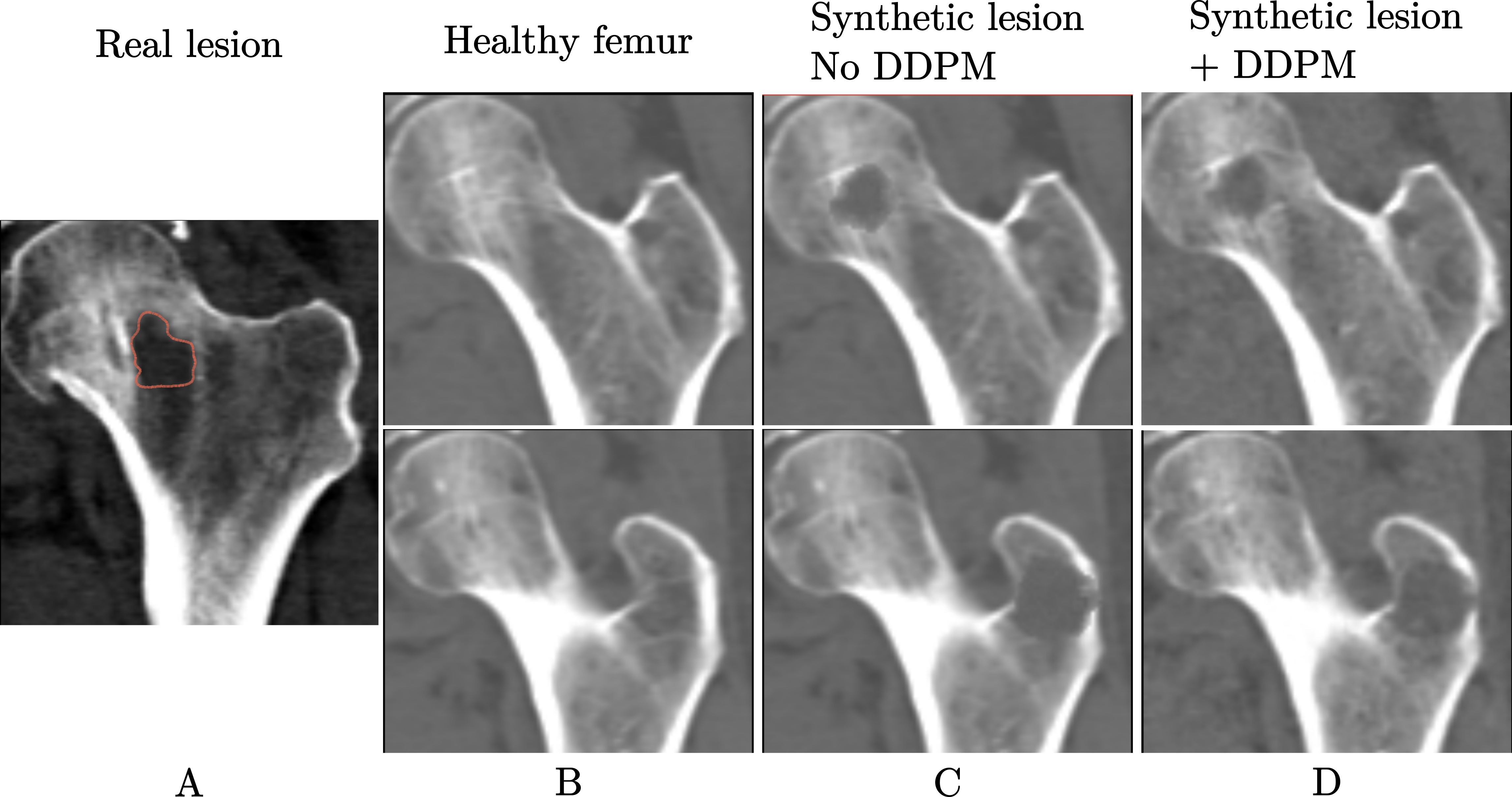}
    \caption{\textsf{Example of data obtained with our synthesis pipeline. Real metastatic volume (A) with its associated segmentation (in red) is used in conjunction with healthy femurs (B) to create two distinct new pathological volumes (C). After using the Denoising Diffusion Probabilistic Model (DDPM), the volumes (D) are obtained.}}
    \label{fig:synthesis_data_examples}
\end{figure}

\subsubsection{Diffusion Network}
\label{ssec:diffusion_network}
We propose a novel approach to enhance the diversity and realism of our synthetic dataset using diffusion models.
We train a custom 3D DDPM \cite{ho_ddpm_2020,kazerouni2023diffusion} to improve the quality of the previously generated synthetic sets. 
Unlike traditional image generation, our focus is not on creating new samples from scratch but on modifying existing images, as inpainting \cite{lugmayr2022repaint} or anomaly detection methods \cite{wyatt2022anoddpm}. 
This enables us to train our model effectively with a smaller training dataset.

Using the following forward (diffusion) process $q$ on the data $x$, iteratively adding gaussian noise following a normal law $\mathcal{N}$, as in \cite{ho_ddpm_2020}: 
\[
q(x_{1:T} |x_0) :=\prod_{t=1}^{T}q(x_t|x_{t-1}), \quad \quad  q(x_t|x_{t-1}) := \mathcal{N} (x_t; \sqrt{1 - \beta_t} x_{t-1}, \beta_t \mathbf{I})  \quad (1)
\]
we choose a variance schedule ranging linearly from  $\beta_1=10^{-4}$ to $\beta_T=2 \times 10^{-3}$, as we only need a narrow range of Gaussian noise for the diffusion process in our application.
This also allows us to train our DDPM using fewer training steps (200).
We trained a 3D DDPM on our 26 healthy femur volumes, and used the trained model on inference on our 5675 synthetic data.
For the sampling stage, we use the Denoising Diffusion Implicit Model (DDIM) \cite{song_ddim_2022} approach to reduce the inference time as much as possible.

\begin{figure}[!tp]
\centering
    \includegraphics[width=0.9\columnwidth, keepaspectratio]{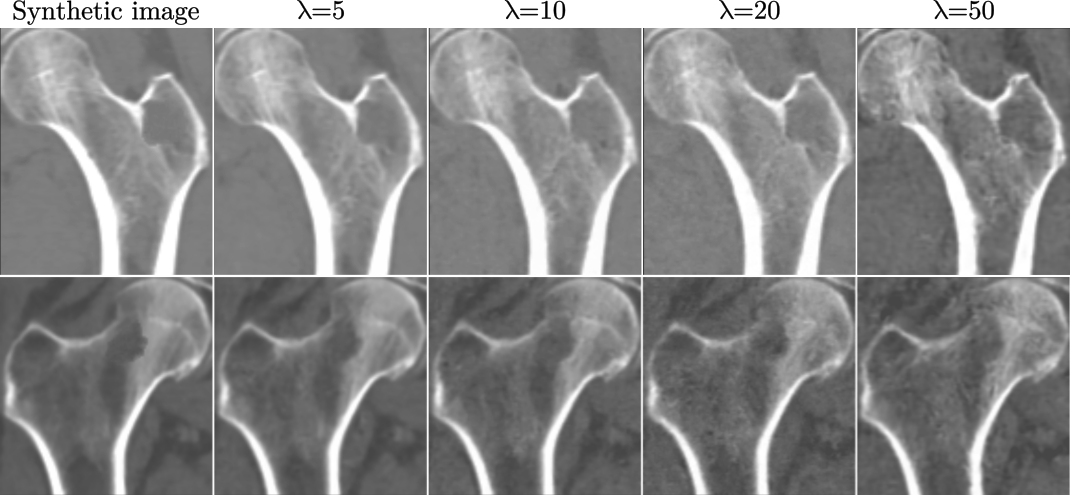}
    \caption{\textsf{Examples of images obtained from two synthetic volumes with artificial lesion located in the greater trochanter on the top row and in the femoral neck on the bottom row after sampling for different values of timesteps $\lambda$.}}
    \label{fig:timesteps}
\end{figure}

Using DDIM, we can reduce the inference steps from 200 to 50 and therefore decrease the computational cost, while maintaining a good overall sampling quality.

After adding noise to a synthetic data $x_0$ until a timestep $\lambda$ ($x_0 \rightarrow x_\lambda$) as shown below, the volume is denoised to $x_0$.
With $\alpha_t = 1- \beta_t$ and $\overline{\alpha_t} = \prod_{i=0}^{T}\alpha_i$ as in \cite{ho_ddpm_2020}:
\[
x_\lambda = x_0 \sqrt{\overline{\alpha_t}} + \epsilon_\lambda \sqrt{1-\overline{\alpha_t}}, \quad \epsilon_\lambda \sim \mathcal{N} (0, \mathbf{I})  \quad (2)
\]
Figure \ref{fig:timesteps} illustrates samples obtained for various timesteps $\lambda$ ranging from 5 to 50.
Fewer timesteps result in less noise and fewer anatomical modifications but limit sample diversity.
Conversely, more timesteps offer greater diversity but may alter anatomical structures, particularly metastasis.
We selected $\lambda=10$ as a balance, where the location of the synthetic lesions still matches the previously generated segmentation.
We trained the 3D DDPM network proposed in version 1.3.0 of the MONAI framework \cite{cardoso_monai_2022} on the 26 healthy femurs at our disposal using an NVIDIA RTX A5000 GPU card (24GB GPU RAM).
To fit our hardware constraints, we modify the denoising UNet used in the reverse process to a 3-level UNet with 128, 256, and 512 channels and an attention mechanism of 512 features only at the deepest level.
The input is also modified to accept $96 \times 96 \times 48$ patches.
The DDPM is optimized with Adam (initial learning rate of $lr=10^{-4}$, exponential decay of 0.999) over 200 timesteps and a batch size of 1.
Using this setup, one week is required to perform a 5000 epochs training.
The DDIM sampling is then performed on the 5675 synthetic images with added noise and $\lambda=10$.
It takes less than a minute to generate a volume using our setup.
This second set of images is named "Diffusion" in the following.

\subsection{Metastasis Segmentation}

The same MONAI framework is used to implement the 3D UNet segmentation network.

For all trainings, the same architecture and hyper-parameters, as the same test folds, are used.
The network consists of a 5-level UNet with 32, 64, 128, 256, and 512 channels.
The input size of patches is fixed to $96 \times 96 \times 64$. 
The DICE loss is optimized with SGD momentum using a weight of 0.9 and an initial learning rate of 0.025, an exponential decay of 0.999 for real data, and 0.99 for both synthetic and diffusion sets.
Classical data augmentation is performed: flips, rotations, scaling, Gaussian noise addition, and contrast enhancement.
An early-stopping criterion is used to stop the trainings with a patience of 200 epochs. 
These trainings are performed on NVIDIA P100 GPUs (16GB GPU RAM) and require 1 day/fold for the real images set, and 6 days/fold for the whole synthetic and diffusion sets. 
Post-processing based on morphological operations is added to improve segmentation performance. 
Fine-tuning on real images uses almost the same parameters except $lr=0.001$, no decay, and a patience of 25 epochs.

Five UNet models are obtained: one trained using real images only (Real), one using synthetic set only (Synthetic), one fine-tuning the previous using real images (Synthetic+FT), one using diffusion set only (Diffusion), one fine-tuning the previous using real images (Diffusion+FT).

We investigate the influence of the number of training samples on the overall segmentation performance, by training additional models with 100, 500 and 2000 synthetic samples respectively, with training time ranging from 12 hours to 3 days.

All models are assessed on our 29 pathological real data using a 5-fold cross-validation. The 9 additional pathological volumes used for the operator variability, which are not included in training, are used in testing, for a total of 38 test samples.
The lesions involved in the tested fold are excluded from any contribution in the synthetic images involved in the remaining training folds.

We quantify the segmentation performance using the DICE score, the Hausdorff distance ($HD$ and $HD_{95}$), and the Average Symmetric Surface Distance (ASSD) \cite{maier_hein_metrics-reloaded_2024}.

\subsection{Statistical analysis}
The statistical significance of segmentation approaches is tested against the results obtained with the network trained using only the real images.
As some variables were not normally distributed even after transformation, we used non-parametric tests.
After running the Kruskal-Wallis test showing differences ($P<0.05$) between the 5 groups (Real, Synthetic, Synthetic+FT, Diffusion, Diffusion+FT), a paired t-test (U Mann-Whitney signed rank test) was used to compare groups 2 by 2.

\section{Results} 
\label{sec:results}

\begin{table}[tp]
\centering
\caption{Inter- and intra-operator variability between novice and expert lesions segmentations. Comparison with our best automatic segmentation model trained on diffusion set with fine tuning on real data.}
\label{tab:operator_variability}
\begin{tabular}{@{}cc|cc|cc|c@{}}
\toprule
  & & \multicolumn{2}{c|}{Inter-operator}           & \multicolumn{2}{c|}{Automatic vs Operators}         & Intra-operator \\ 
\cmidrule{2-7}
     & & Novice/Novice & Novice/Expert & vs Novices &  vs Expert & Novices                                                                                               \\
\midrule
Mean DICE    & & $0.73 \scriptstyle{\pm 0.23}$ & $0.72 \scriptstyle{\pm 0.16}$ & $0.71 \scriptstyle{\pm 0.18}$ &  $0.75 \scriptstyle{\pm 0.14}$   & $0.77 \scriptstyle{\pm 0.18}$ \\  
\bottomrule
\end{tabular}
\end{table}

The results on lesion segmentation operator variability are displayed in Table \ref{tab:operator_variability}.
Both the intra- and inter-operator variability are substantial, with an intra-operator DICE of 0.77 and an inter-operator DICE of 0.73 between novices and 0.72 between expert and novice. Those results are comparable to what can be found in existing studies \cite{ataei_effect_2024,rachmil2024automatic}.
Automatic segmentations reach a mean DICE of 0.71 with novice operators and 0.75 with the expert. 

The visual quality of the synthetic images depicted in Figure \ref{fig:synthesis_data_examples} is notable, although, in some instances, the added lesion may appear insufficiently integrated into the image.
Nevertheless, after only a few steps of the diffusion process, the integration of the lesion becomes more realistic. 

The results obtained depending on the amount of synthetic data can be observed in Table \ref{tab:segmentation_different_set_size}. 
As expected, adding more synthetic data improves the segmentation performance. 
However, we can observe that performance reaches a plateau, with comparable results obtained when using 2000 or 5975 samples.
While HD, HD95 and ASSD keep decreasing when using more samples, the DICE stops improving, with even a slight decrease.

\begin{figure}[tp]
\centering
    \includegraphics[width=\columnwidth, keepaspectratio]{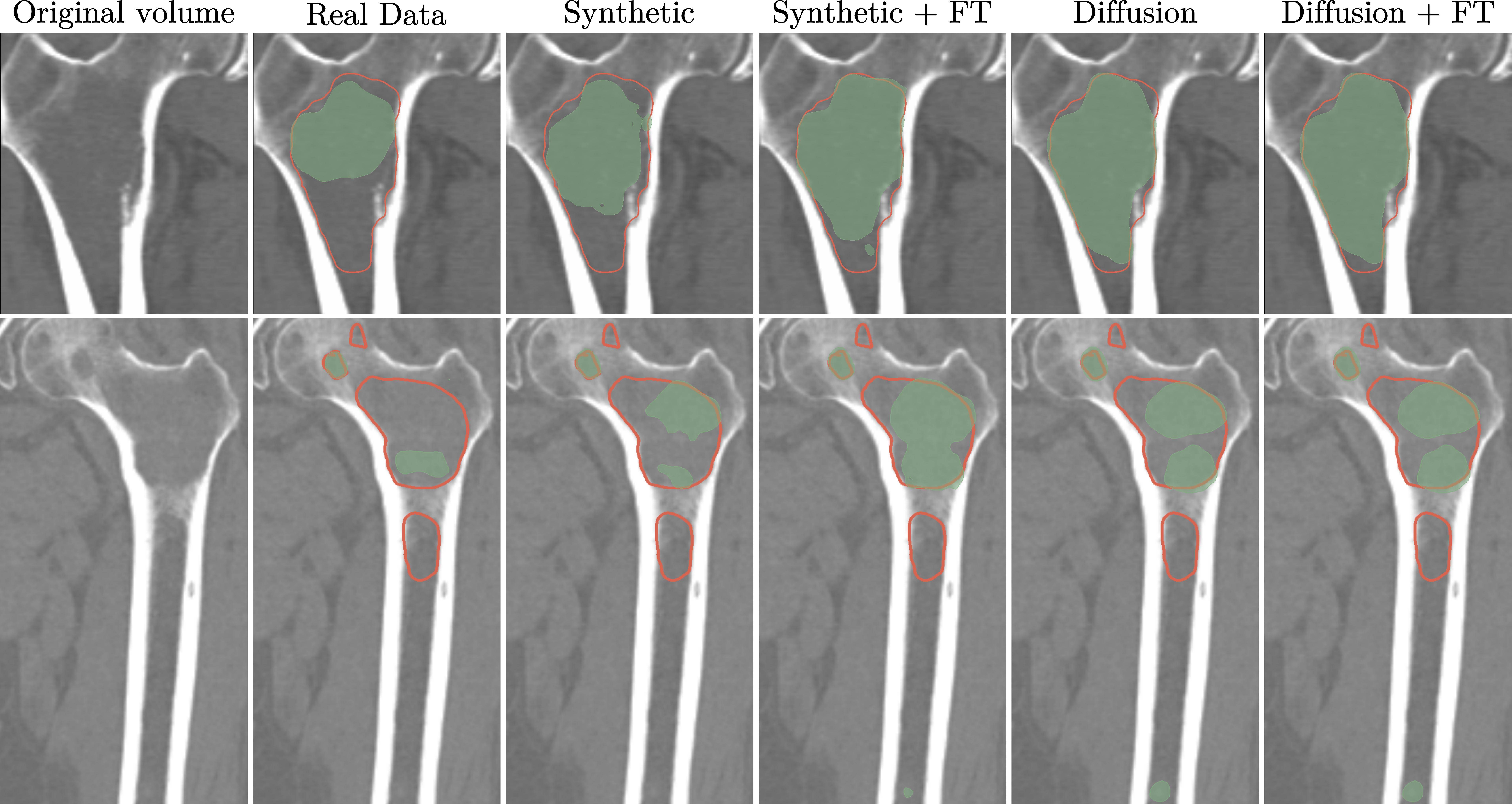}
    \caption{\textsf{Examples of segmentations of two real CT-scans of femurs with metastasis using the five trained models. The ground truth is in red, the automatic segmentations are in green. One can note the large mistakes made by the networks on the lower end of the diaphysis on the three last columns.}}
    \label{fig:segmentations_autos_examples}
\end{figure}

Figure \ref{fig:segmentations_autos_examples} provides a comparison with real metastasis and showcases the automatic segmentations obtained using the five differently trained models.

The contribution of synthetic data with or without diffusion and fine-tuning is apparent, with clear metrics improvement compared to the model trained on real data only.
Some lesions are not accurately segmented and the algorithms can also produce false positives that can negatively impact metrics such as Hausdorff values.

The distance metrics show important differences between the models, with the best model in terms of segmentation performance (DC=0.65) not having the best results in terms of HD and ASSD.
This indicates a presence of false positives that can be further away from the ground-truth segmentation, but given the very close values obtained on ASSD and the better DICE value, the model "Diffusion + FT" is the best overall.
Improving the dedicated post-processing could help diminish the differences between the models in terms of HD.

\begin{table}[tp]
\centering
\caption{Segmentation results on the same test set using four metrics (average $\pm$ std, $HD$, $HD_{95}$ and $ASSD$ are in mm), depending on the training size of the synthetic set. Bold values represent the best score for a given metric, and arrow direction indicates a metric improvement.}
\label{tab:segmentation_different_set_size}
\begin{tabular}{@{}cc|cccc@{}}
\toprule
Training set size  & & $DICE \uparrow $            & $HD \downarrow$          & $HD_{95}\downarrow$ & $ASSD \downarrow$ \\ 
\midrule
100  & & $0.19 \scriptstyle{\pm 0.27}$ & $125.5 \scriptstyle{\pm 133.6}$ & $117.8 \scriptstyle{\pm 130}$ &   $ 58.8\scriptstyle{\pm 89.4}$   \\
\midrule
500  & & $0.31 \scriptstyle{\pm 0.36}$ & $89.4 \scriptstyle{\pm 98.7}$ & $85.0 \scriptstyle{\pm 98.9}$ &  $30.2 \scriptstyle{\pm 55.6}$   \\
\midrule
2000 & & $\mathbf {0.62}\scriptstyle{\pm 0.29}$ & $79.8 \scriptstyle{\pm 101.9}$ & $66.8 \scriptstyle{\pm 92.1}$ &  $12.8 \scriptstyle{\pm 34.3}$   \\
\midrule
5675 (full)    & & $0.59 \scriptstyle{\pm 0.29}$ & $\mathbf {50.7} \scriptstyle{\pm 61.4}$ & $\mathbf {45.5} \scriptstyle{\pm 62.2}$ &  $\mathbf{11.5}  \scriptstyle{\pm 30.2}$   \\
\bottomrule
\end{tabular}
\end{table}

These observations are confirmed in Table \ref{tab:segmentation_results} displaying the four metrics statistics for the five trained networks: $HD$, $HD_{95}$ and $ASSD$ are large and with an important standard deviation.
The DICE score improves from 0.45 to 0.59 using the synthetic images, and 0.62 when diffusion images are used.

The fine-tuning step improves the segmentation quality, although only slightly, with a DICE reaching 0.65 when diffusion images are used alongside fine-tuning.

However, for the four metrics, there was only a significant difference for the DICE score using the non-parametric Kruskal-Wallis test.
There was no significant difference between the four tested algorithms (Synthetic, Synthetic+FT, Diffusion, Diffusion+FT) in terms of DICE (U Mann-Whitney). 
Only when comparing the tested algorithms with the real data, the DICE parameter shows a significant difference.

\begin{table}[tp]
\centering
\caption{Segmentation results on the same test set using four metrics (average $\pm$ std, $HD$, $HD_{95}$ and $ASSD$ are in mm) with the 3DUNet optimized using three training sets (real images only, synthetic images only, and diffusion synthetic images only) followed or not by a fine-tuning on real images.
Bold values represent the best score for a given metric, and arrow direction indicates a metric improvement.} 
\label{tab:segmentation_results}
\begin{tabular}{@{}cc|cccc@{}}
\toprule
Training set  & & $DICE \uparrow $            & $HD \downarrow$          & $HD_{95}\downarrow$ & $ASSD \downarrow$ \\ 
\midrule
Real         & & $0.45 \scriptstyle{\pm  0.33}$ & $59.8 \scriptstyle{\pm 67.3}$ & $54.9 \scriptstyle{\pm 67.4}$ &   $ 20.0 \scriptstyle{\pm 43.1}$   \\
\midrule
Synthetic    & & $0.59 \scriptstyle{\pm 0.29}$ & $\mathbf {50.7} \scriptstyle{\pm 61.4}$ & $\mathbf {45.5} \scriptstyle{\pm 62.2}$ &  $\mathbf{11.5}  \scriptstyle{\pm 30.2}$   \\
Synthetic+FT & & $0.62 \scriptstyle{\pm 0.27}$ & $58.4 \scriptstyle{\pm 70.2}$ & $53.3 \scriptstyle{\pm 70.5}$ &  $11.9 \scriptstyle{\pm 28.7}$   \\
\midrule
Diffusion    & & $0.62 \scriptstyle{\pm 0.27}$ & $87.7 \scriptstyle{\pm 104.6}$ & $78.8 \scriptstyle{\pm 97.6}$ &  $16.5 \scriptstyle{\pm 31.2}$   \\
Diffusion+FT & & $\mathbf{0.65} \scriptstyle{\pm 0.24}$ & $72.3\scriptstyle{\pm 97.4}$ & $62.9 \scriptstyle{\pm 92.6}$ &  $12.4 \scriptstyle{\pm 22.8}$   \\      
\bottomrule
\end{tabular}
\end{table}

\section{Discussion} 
\label{sec:discussion}

We have generated synthetic volumes to alleviate the issue of data scarcity for our segmentation task.

We observed that most synthesized volumes have an overall convincing visual quality and although some samples do not have a high realism, our results show that they still contribute positively to the training and thus improve segmentation performance.

The results obtained depending on the synthetic training dataset size show that with larger amounts of training data, a better segmentation performance is achieved by the model up until a limit which could be estimated more precisely.
A trade-off exists between the amount of synthetic data used in training and the total training time.
In addition, the more data is added, the more stable the training becomes.
We therefore opted to use the synthetic training dataset with the largest amount of samples to improve segmentation quality.
Additional testing could allow to optimize the number of training samples used considering constraints on GPU resources.

Segmentation of bone metastasis remains a major challenge for both expert and automated segmentation approaches.
The current state of the art on a similarly sized dataset achieves a maximum DICE score close to 0.5\cite{ataei_effect_2024,rachmil2024automatic}, which is coherent with our results obtained with our model trained on real data.
Our models achieve a best DICE of 0.65, which is better than existing models even when those are trained on a significantly bigger dataset.
Our proposed pipeline demonstrates superior performance compared to previous studies, despite using less annotated data, showcasing its efficiency. 

Inter-operator segmentation variability is very high in our dataset (DC$<$0.75), underlining the difficulty of the task.
On the data used for the operator variability study, the automatic segmentations achieve a DICE comparable to inter-operator variability, and performed better than the novice segmentations when compared to expert segmentation, which goes towards validating the use of automatic models for bone metastasis segmentation. 
However, we can note that the automatic model performed better on this dataset than on the one it was previously trained and evaluated. 
This further indicates that the performance is linked to the dataset used, and a more in-depth evaluation could be done to better judge the models' performance.
Intra-operator variability is also substantial for novice annotators, which shows the difficulty of precisely defining the borders of the metastasis when manually annotating.

The primary enhancement of our approach for segmentation lies in the increased data volume generated through the synthesis process, rather than traditional image enhancement techniques during training. 
While the contribution of DDPM/DDIM on metrics is subtle, it still globally improves the quality of the segmentation and enables the synthesis of more realistic images.
Fine-tuning offers minimal contribution but has potential for optimization. 

However, notable segmentation errors persist, and comprehensive testing on datasets from multiple centers is essential to ascertain generalizability. 
Despite these challenges, the simplicity of our approach could allow each individual center to customize their networks, potentially achieving comparable performance with minor adjustments fitting reasonable hardware constraints.

\section{Conclusion}
\label{sec:conclusion}
We propose a data synthesis pipeline that produces realistic and varied metastatic images, allowing to better train lesions segmentation models and thus achieve better performance than comparable state-of-the art models.

Our data synthesis pipeline enables the generation of a large quantity of volumes along with ground-truth labels, substantially enhancing overall segmentation performance. 
Incorporating 3D diffusion models enhances the realism and diversity of synthetic images, although further investigation into the generated samples is necessary to comprehensively assess model performance.
The segmentation results obtained with our approach are encouraging, but challenges remain in addressing segmentation and detection errors and extending testing across diverse datasets to enable clinical applications.

\section*{Acknowledgments}

The authors thank master student Elise Jourdain who served as segmentation beginner for the variability study.
They also thank medical student Stéphane Cadot for his help on metastasis segmentation.
This work was partly funded by LabEx Primes, France (ANR-11-LABX-0063) and MSDAvenir Research Grant. The images were selected and acquired in clinical routine thanks to the MEKANOS clinical research assistants. 

%
%
\clearpage
\bibliographystyle{splncs04}
\bibliography{references.bib}

\begin{thebibliography}{10}
\providecommand{\url}[1]{\texttt{#1}}
\providecommand{\urlprefix}{URL }
\providecommand{\doi}[1]{https://doi.org/#1}

\bibitem{cardoso_monai_2022}
Monai: An open-source framework for deep learning in healthcare (Nov 2022). \doi{10.48550/arXiv.2211.02701}

\bibitem{ataei_evaluation_2021}
Ataei, A., Eggermont, F., Baars, M., van~der Linden, Y., de~Rooy, J., Verdonschot, N., Tanck, E.: Evaluation of inter- and intra-operator reliability of manual segmentation of femoral metastatic lesions. International Journal of Computer Assisted Radiology and Surgery  \textbf{16}(10),  1841--1849 (Oct 2021). \doi{10.1007/s11548-021-02450-w}

\bibitem{ataei_effect_2024}
Ataei, A., Eggermont, F., Verdonschot, N., Lessmann, N., Tanck, E.: The effect of deep learning-based lesion segmentation on failure load calculations of metastatic femurs using finite element analysis. Bone  \textbf{179},  116987 (Feb 2024). \doi{10.1016/j.bone.2023.116987}

\bibitem{fernandez2022can}
Fernandez, V., Pinaya, W.H.L., Borges, P., Tudosiu, P.D., Graham, M.S., Vercauteren, T., Cardoso, M.J.: Can segmentation models be trained with fully synthetically generated data? In: International Workshop on Simulation and Synthesis in Medical Imaging. pp. 79--90. Springer (2022)

\bibitem{gardegaront2024inter}
Gardegaront, M., Sas, A., Brizard, D., Levillain, A., Bermond, F., Confavreux, C.B., Pialat, J.B., van Lenthe, G.H., Follet, H., Mitton, D.: Inter-laboratory reproduction and sensitivity study of a finite element model to quantify human femur failure load: case of metastases. Journal of the Mechanical Behavior of Biomedical Materials p. 106676 (2024)

\bibitem{goodfellow_GAN_2014}
Goodfellow, I., Pouget-Abadie, J., Mirza, M., Xu, B., Warde-Farley, D., Ozair, S., Courville, A., Bengio, Y.: Generative adversarial nets. In: Ghahramani, Z., Welling, M., Cortes, C., Lawrence, N., Weinberger, K.Q. (eds.) Advances in {Neural} {Information} {Processing} {Systems}. vol.~27. Curran Associates, Inc. (2014)

\bibitem{ho_ddpm_2020}
Ho, J., Jain, A., Abbeel, P.: Denoising diffusion probabilistic models (Dec 2020), \url{http://arxiv.org/abs/2006.11239}

\bibitem{huo_deep_2023}
Huo, T., Xie, Y., Fang, Y., Wang, Z., Liu, P., Duan, Y., Zhang, J., Wang, H., Xue, M., Liu, S., Ye, Z.: Deep learning-based algorithm improves radiologists’ performance in lung cancer bone metastases detection on computed tomography. Frontiers in Oncology  \textbf{13},  1125637 (Feb 2023). \doi{10.3389/fonc.2023.1125637}, \url{https://www.ncbi.nlm.nih.gov/pmc/articles/PMC9946454/}

\bibitem{kazerouni2023diffusion}
Kazerouni, A., Aghdam, E.K., Heidari, M., Azad, R., Fayyaz, M., Hacihaliloglu, I., Merhof, D.: Diffusion models in medical imaging: A comprehensive survey. Medical Image Analysis p. 102846 (2023)

\bibitem{lugmayr2022repaint}
Lugmayr, A., Danelljan, M., Romero, A., Yu, F., Timofte, R., Van~Gool, L.: Repaint: Inpainting using denoising diffusion probabilistic models. In: Proceedings of the IEEE/CVF Conference on Computer Vision and Pattern Recognition. pp. 11461--11471 (2022)

\bibitem{macedo2017}
Macedo, F., Ladeira, K., Pinho, F., Saraiva, N., Bonito, N., Pinto, L., Goncalves, F.: Bone metastases: An overview. Oncology Reviews  \textbf{11}(1) (Mar 2017). \doi{10.4081/oncol.2017.321}, publisher: Frontiers Media SA

\bibitem{maier_hein_metrics-reloaded_2024}
Maier-Hein, L., Reinke, A., Godau, P., Tizabi, M.D., Buettner, F., Christodoulou, E., Glocker, B., Isensee, F., Kleesiek, J., Kozubek, M., et~al.: Metrics reloaded: recommendations for image analysis validation. Nature Methods  \textbf{21}(2),  195–212 (Feb 2024). \doi{10.1038/s41592-023-02151-z}

\bibitem{paranavithana_systematic_2023}
Paranavithana, I.R., Stirling, D., Ros, M., Field, M.: Systematic {Review} of {Tumor} {Segmentation} {Strategies} for {Bone} {Metastases}. Cancers  \textbf{15}(6), ~1750 (Jan 2023). \doi{10.3390/cancers15061750}, \url{https://www.mdpi.com/2072-6694/15/6/1750}, number: 6 Publisher: Multidisciplinary Digital Publishing Institute

\bibitem{rachmil2024automatic}
Rachmil, O., Artzi, M., Iluz, M., Druckmann, I., Yosibash, Z., Sternheim, A.: Automatic segmentation of femoral tumors by nnu-net. Clinical Biomechanics  \textbf{116},  106265 (2024)

\bibitem{ronneberger2015}
Ronneberger, O., Fischer, P., Brox, T.: U-net: Convolutional networks for biomedical image segmentation. In: Medical Image Computing and Computer-Assisted Intervention--MICCAI 2015: 18th International Conference, Munich, Germany, October 5-9, 2015, Proceedings, Part III 18. pp. 234--241. Springer (2015)

\bibitem{saillard2024finite}
Saillard, E., Gardegaront, M., Levillain, A., Bermond, F., Mitton, D., Pialat, J.B., Confavreux, C., Grenier, T., Follet, H.: Finite element models with automatic computed tomography bone segmentation for failure load computation. Scientific Reports  \textbf{14}(1),  16576 (2024)

\bibitem{song_ddim_2022}
Song, J., Meng, C., Ermon, S.: Denoising {Diffusion} {Implicit} {Models} (Oct 2022), \url{http://arxiv.org/abs/2010.02502}

\bibitem{wyatt2022anoddpm}
Wyatt, J., Leach, A., Schmon, S.M., Willcocks, C.G.: Anoddpm: Anomaly detection with denoising diffusion probabilistic models using simplex noise. In: Proceedings of the IEEE/CVF Conference on Computer Vision and Pattern Recognition. pp. 650--656 (2022)

\end{thebibliography}

\end{document}